# Modèles de coûts en fonderie sable : les limites d'une approche générique

N.PERRY, M.MAUCHAND, A.BERNARD

IRCCyN

*Eq. Proj. Ingénierie Virtuelle pour le Génie Industriel*
*IRCCyN (UMR CNRS 6597) / Ecole Centrale de Nantes*
*1, rue de la Noë - B.P. 92101*
*44321 Nantes CEDEX 03, France*
*Nicolas.Perry@irccyn.ec-nantes.fr*

RÉSUMÉ. *La maîtrise des coûts, au plus tôt du cycle de vie produit, est devenue un atout majeur dans la compétitivité des entreprises confrontées à la mondialisation de la concurrence. Après avoir mis en avant les problématiques liées à la difficulté de cette maîtrise, nous présenterons une approche définissant un concept d'entité coût lié au domaine d'activité du produit à concevoir et réaliser. Nous chercherons alors à appliquer cette approche aux domaines de la fonderie sable. Ce travail mettra en avant les difficultés de hiérarchisation des entités composant les modèles ainsi créés ainsi que les limites d'une approche générique.*

ABSTRACT. *The control of the costs, as soon as possible of the product life cycle, became a major asset in the competitiveness of the companies confronted with the universalization of competition. After having proposed the problems related this control difficulties, we will present an approach defining a concept of cost entity related to the activities of the product to be designed and realized. We will then try to apply this approach to the fields of the sand casting. This work will highlight the hierarchisation difficulties with the entities composing the models created as well as the limits of a generic approach.*

MOTS-CLÉS: *management à coûts objectifs, modèle de produit, modèle de coût, entité coût, management par les activités.*

## 1. Introduction

Dès les années 70-80, des rapports et études concernant le rôle stratégique des activités de conception ont été édités en Grande Bretagne et aux Etats-Unis, en vue de convaincre les entreprises et les pouvoirs publics de s'orienter vers une nouvelle approche afin d'améliorer les performances économiques industrielles. Le rôle primordial de la qualité de la conception a été renforcé aux Etats-Unis par le rapport Made in America de la Commission sur la productivité du MIT. Ce rapport met en avant le fait que l'activité de conception possède un impact fondamental dans le processus d'innovation et dans la concurrence entre entreprises et nations. Ces conclusions étaient confirmées par le rapport Improving Engineering Design, Designing for Competitive Advantage (1991) établi par le comité « Engineering Design Theory and Methodology » du National Research Council des Etats-Unis. La conception est le facteur clé du processus de développement de produit. La compétitivité d'une nation, dans le contexte mondial actuel passe par la capacité à développer de nouveaux produits de haute qualité, de faible coût et qui correspondent aux demandes des utilisateurs [Perrin 1996]. Nous rappellerons ici les facteurs contextuels qui ont conduit à cette mise en avant du coût et de la nécessité de sa maîtrise au plus tôt en conception et développement de produits. Dans cette optique, nous avons mené, en partenariat avec la société Cognition Europe, une étude sur les modèles de coûts en fonderie sable, se basant sur le concept d'entité coûts. Ce travail avait pour objectif de proposer une approche pour calculer les coûts, en phase de conception/développement de produit. L'objectif de cette communication est d'en exposer la teneur en tant qu'approche générique, et d'en argumenter les limites.

## 2. Evolution du contexte industriel

Le contexte économique a fortement évolué ces dernières décennies (concurrence internationale, globalisation des marchés…). « Le paradis des producteurs est devenu celui du client » [Bouin & Simon 2000]. En parallèle, le client a aussi évolué ; il exige des produits innovants, de qualité, réalisés dans des délais plus courts avec des services associés adaptés à ses besoins et, bien évidemment, le tout à des prix de plus en plus réduits. Ainsi, le client définit non seulement les exigences techniques des produits, mais il impose aussi le prix (prix du marché). Ainsi, pour répondre à ces nouvelles contraintes et s'adapter, les entreprises doivent faire preuve d'une réactivité et d'une agilité accrues pour rester compétitives.

D'autre part, le champ concurrentiel est devenu mondial. Ceci a sensibilisé les entreprises sur les aspects décisionnels à savoir, réduire les temps de prise de décision (disponibilité d'informations pertinentes au bon moment) et réduire les risques de mauvaises décisions (visibilité complète des impacts de celles-ci sur la totalité du cycle de vie produit / processus). C'est pourquoi il est indispensable de disposer d'indicateurs fiables permettant de quantifier a priori les coûts et délais résultant d'une conception ou d'une innovation sur un produit à un moment quelconque du cycle de vie. De tels indicateurs intègrent conjointement des points de vues techniques et économiques tout au long du cycle de vie produit. A l'heure actuelle de tels indicateurs n'existent pas, ou bien donnent une perception incomplète de la situation, ou encore manquent de fiabilité. En 1998, ces constats ont préludé à la formulation du projet METACOG[1], (Méthodologie de conception à coût objectif global), qui avait pour objectif de définir ces indicateurs en unifiant les compétences et points de vues complémentaires de partenaires tels que RENAULT, le LAMIH (équipe de recherche sur les méthodes de conception et l'évaluation des performances des systèmes complexes de production) et le LRSGUN (équipe de recherche en sciences de gestion).

Le calcul des coûts connaît depuis le dernier quart du 20ème siècle une évolution majeure. Alors que depuis ses débuts, le calcul des coûts est utilisé pour améliorer la connaissance des coûts des produits en production, son application s'est progressivement déplacée vers le coût des produits futurs. Ce changement d'objet ne ressort pas d'une évolution de la technique de calcul de coûts mais d'un changement dans la compréhension et dans la modélisation des coûts. Le passage au cycle de vie apparaît comme une extension, sur un horizon plus long, du mouvement amorcé dès les années 70. « Les outils de gestion, comme l'ensemble des autres techniques, sont apparus, le plus souvent, afin d'apporter des solutions aux problèmes de leur époque. Leur construction est contingente ; elle tient compte, naturellement, de l'environnement économique, de la structure et du fonctionnement de l'entreprise. » [Milkoff 1996]

## 3. Maîtrise de la valeur – Maîtrise des coûts

La notion de coût ou de valeur (d'un produit ou d'un service) a fortement évolué elle aussi avec le contexte. Aujourd'hui, le prix de vente dépend du prix que le client est prêt à payer en fonction des facteurs de valeur qu'il apprécie (le prix de vente est guidé par le marché). En conséquence, la marge ne sert plus à calculer le prix de vente, il en résulte. Ainsi, la marge et le prix de vente doivent être des objectifs ; ils permettent de déterminer un « prix de revient maximum objectif », dit encore « coût objectif ».

La rentabilité, elle aussi, est souvent fondée sur des critères dépassés, estimée par le rapport de la marge sur le prix de vente. Or, des pièges se cachent derrière cette apparente logique. D'une part, le concept de prix de revient complet, et beaucoup plus que le coût d'un produit, d'autre part, il est aberrant de comparer une marge au prix de vente [Bernard 2003, Brodier 1990].

---

[1] http://www.univ-valenciennes.fr/PROSPER/serveur/doc/lyon/METACOG.doc

Selon Perrin les indicateurs portant sur les coûts sont à classer principalement parmi les indicateurs de résultats[Perrin 1996]. Les deux critères considérés sont d'un côté le coût du produit et de l'autre le respect du budget.

- Le respect du coût du produit est notamment mis en avant dans le cadre du développement des méthodes de target costing (gestion par coût-cible) et ABC (Activity Based Costing). L'idée est de fixer un coût cible comme objectif auquel on doit faire correspondre le coût de revient du produit. L'indicateur alors utilisé est le rapport entre le coût de revient effectif du produit et le coût objectif.

- Le respect des budgets, en particulier des budgets d'études est le second élément à prendre en compte au niveau financier. L'indicateur est alors naturellement le rapport du montant du dépassement de budget sur le budget initialement fixé.

Les évolutions contextuelles présentées précédemment impactent la nature des coûts dans l'entreprise. Autrefois, les coûts de revient comprenaient une grande majorité de coûts directs (souvent de l'ordre de 70%), c'est-à-dire directement affectés aux produits. Les autres coûts pouvaient faire l'objet de répartitions globales ; le choix de la clé de répartition influait peu sur le résultat. C'est la méthode de la comptabilité analytique (cf. Fig.1) avec une répartition des charges indirectes sous forme de pourcentage fixe d'un élément direct (souvent le nombre d'heures directes), ce qui peut s'admettre tant que la proportion de charges indirectes n'est pas trop forte.

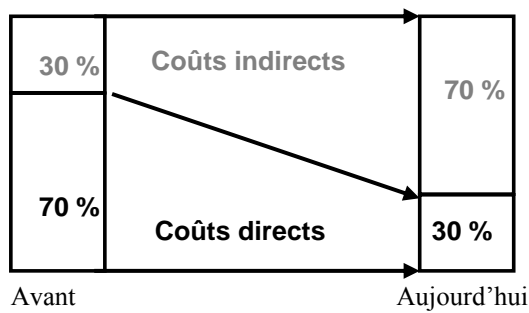

**Figure 1.** *Evolution de la répartition des coûts directs et indirects*

Il n'en va plus ainsi actuellement où les proportions ont été inversées, les coûts directs ne constituent souvent plus que 30 % du coût total, comme l'indique la figure 2. Non seulement la répartition des charges indirectes, en fonction de clés globales, crée un facteur arbitraire dans le calcul des coûts de revient, mais elle obscurcit le contrôle.

Ces deux préoccupations ont donné lieu à de nombreuses recherches qui ont fini par modifier profondément les systèmes de calcul des coûts de revient.

Ainsi, les clés de répartition des charges indirectes d'une part créent un facteur arbitraire, d'autre part obscurcissent le contrôle. Les méthodes et modèles de calcul des coûts doivent donc profondément être modifiés, conduits par le biais de nombreux travaux de recherche.

Reprenons le cycle de vie d'un produit et analysons l'impact des décisions en terme de coûts engendrés. La phase de « préconception-conception » fige entre 70 et 85% du coût du produit et, en fin de conception détaillée la marge de gain sur le coût final est très limitée. A contrario, l'évolution des dépenses réellement engagées par l'entreprise est très limitée dans ces phases amont. Or, c'est là que le contrôle des coûts peut réellement s'exercer.

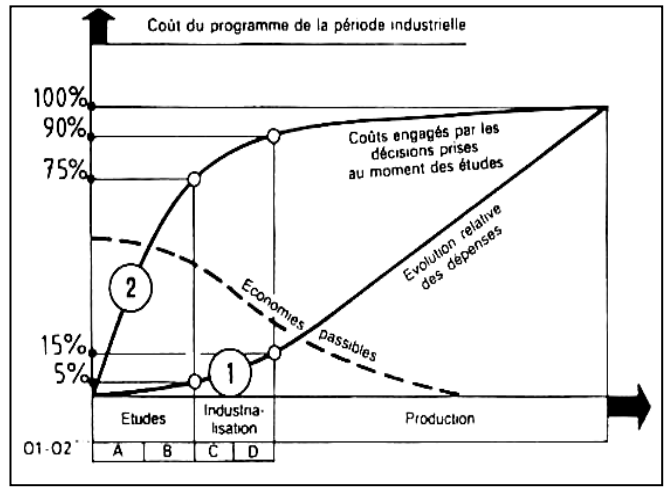

**Figure 2.** *Evolution des coûts au cours du cycle de vie d'un produit [Alcouffe & Bes 1997]*

En conséquence, plus on avance dans la définition du produit, moins il est aisé de réduire les coûts et les modifications sont de plus en plus coûteuses. Il y a donc un intérêt économique évident à se concentrer sur la recherche d'une optimisation du produit le plus en amont possible du processus, les potentialités de gains étant importantes et les coûts engagés faibles.

Ainsi donc selon Bellut, « le coût d'un produit à venir nait au bureau d'études » [Bellut 2002]. Les choix de conception, peu coûteux en termes de consommation de ressources peuvent se révéler très coûteux en fabrication, à un moment où il devient extrêmement difficile de procéder à des modifications. [Mevellec 2000].

Les ratios mentionnés ont pour but de fixer les idées et peuvent évoluer avec le produit, le métier, la série ou le degré d'innovation, mais ne remettent pas en cause le principe.

**4. La conception pour un coût objectif**

Comme l'Analyse de la Valeur, c'est aux Etats-Unis qu'est née dans les années 60 la Conception pour un Coût Objectif désignée par l'expression Design – To – Cost (DTC). C'est une méthode permettant de concevoir un produit en imposant dès le départ un coût plafond [Delafollie 1991]. La Conception à Coût Objectif peut être définie comme « un principe d'action visant à établir des objectifs rigoureux dès le stade de développement d'un système et à autoriser des compromis coût/performance pour permettre le respect de ces objectifs » [Petitdemange 1991]. La contrainte « coût » devient ici capitale et l'on peut dire que cette donnée est à mettre, dans cette perspective, au même niveau que les performances techniques, lors de l'étude d'un produit nouveau. Il est à noter qu'alors le coût objectif souhaité devient la contrainte et que les grandeurs variables sont les performances techniques.

L'intérêt d'intégrer le coût de production dès la conception est partagé par le client et l'industriel puisqu'il permet de maîtriser le développement en fonction des besoins exacts des futurs utilisateurs (il incite à la recherche d'idées nouvelles exigées par les contraintes économiques), d'autre part, il permet de préparer et d'organiser sa production très tôt et de mieux maîtriser sa marge [Giannopoulos 2003].

Cette méthode de conception reprend les démarches d'activités ou de tâches avec mise en parallèle du coût seuil souhaité. Ceci conduit à un rallongement du temps de conception puisque les dépassements successifs obligent à reprendre l'étude de la tâche ou à statuer sur les futurs réductions à réaliser (techniques ou économiques).

Par contre, le pilotage à coût objectif rompt avec la démarche habituelle du pilotage par les coûts en proposant une approche tournée vers le marché. De plus, il conduit à une « transversalisation » des fonctions au sein de l'entreprise et peut être utilisé conjointement avec des méthodes d'analyse de coûts du type ABC. Il permet également une analyse très approfondie des fonctions d'un produit en relation avec ses coûts. Enfin, il insuffle une dynamique forte dans l'entreprise fondée sur une gestion par objectifs et un travail d'équipe de projet [Kieffer 2002].

On peut toutefois faire remarquer que l'allongement du temps d'étude amont contribue à ce que la phase de mise au point (pré-série) ne soit pas perturbée par diverses modifications - voire par une remise en cause du produit - comme cela peut arriver dans une démarche classique. On accepte en fait de "perdre" du temps à la phase conception, donc en amont avec la quasi-certitude de le regagner en aval avant la date de mise sur le marché du produit. A fortiori, le recours à la CCO permet de ne pas avoir à modifier le produit après son installation sur le marché.

**4.1.** *le concept d'entité coûts, logique de modélisation*

C'est dans ce contexte que se place notre étude. L'objet est de réussir à maîtriser plus finement l'ensemble des coûts (directs et indirects) dans la réalisation de pièces mécaniques en fonderie sable. Comme indiqué précédemment, il est impératif de pouvoir donner un outil aux ingénieurs du bureau d'étude dans le but de maîtriser les coûts de conception des pièces. En collaboration avec la société *Cognition Europe*, et sur la base de l'outil *Cost Advantage*, nous travaillons sur les modèles de coûts à appliquer dans le cas de la fonderie sable de pièces en acier.

Nous nous appuyons sur un précédent travail, proposant une approche intégrée pour la fonderie sable, réalisée dans le cadre d'une thèse en partenariat avec l'entreprise SMC Colombier Fontaine (France) du groupe AFE Métal (Bernard 2002). Ce travail, à permis d'une part d'assoire et de formaliser la base de connaissances métier nécessaire à la maîtrise du cycle de vie produit dans une entreprise de fonderie sable. D'autre part, nous avons validé une approche, une méthodologie et un déploiement permettant d'assurer une exacte connaissance des coûts des pièces et de leur impact sur le rendement de l'entreprise (Perry 2003).

Mais commençons par faire un rappel sur le concept d'entité coût et de contexte, base de notre modélisation.

*4.1.1. Entités coûts*

Une Entité Coût est un groupement de coûts associés aux ressources consommées par une activité. La condition générale tient à l'homogénéité des ressources, ce qui permet d'associer un inducteur unique à l'entité coût [H'Mida 2002).

Mais qu'est ce qu'une entité ? C'est un groupement sémantique (atome de modélisation) caractérisé par un ensemble de paramètres, utilisé pour décrire un objet indécomposable utilisé dans le raisonnement relatif à une ou plusieurs activités liées à la conception et l'utilisation des produits et des systèmes de production. Par ce biais, la modélisation permet de formaliser de l'expertise, de capitaliser du savoir-faire, de disposer dès la phase de conception d'informations liées aux activités de réalisation. Enfin, elle améliore la communication entre les intervenants tout au long du cycle de vie du produit.

*4.1.2. Les contextes*

Les contextes viennent instancier et spécifier les entités définies. Elles sont définies à trois niveaux dans notre modélisation (imposée par l'outil). La première est définie à un niveau process, la seconde à un niveau matériau et la dernière est directement lié à l'entité (feature). Ce contexte correspond à un nœud d'arborescence (liant un processus à un matériau et à une entité) représentant l'environnement dans une vue donnée (cf Figure 3). On définit ainsi l'ensemble des contextes nécessaires à la réalisation d'une pièce en fonction de la gamme exacte choisie ou anticipée pour cette pièce (cf. Figure 4).

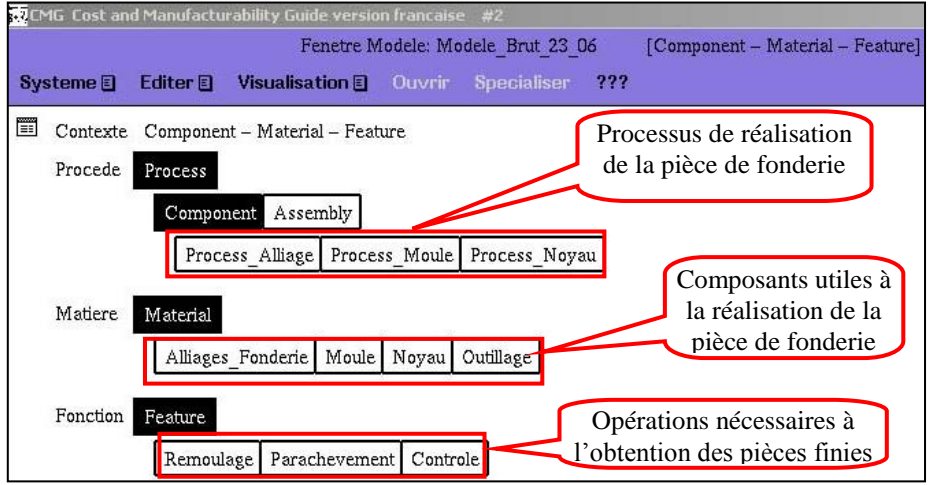

**Figure 3.** *Modélisation en fonderie Sable*

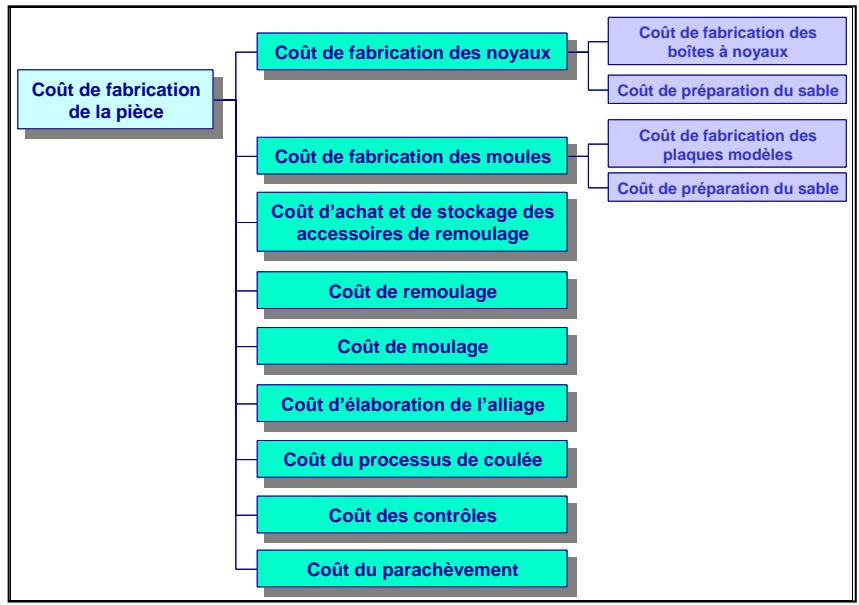

**Figure 4.** *Structure arborescente des coûts, exemple de la fonderie sable*

*4.2. Modélisation en fonderie sable*

A partir de cette analyse et de cette mise en situation, l'objectif est la création d'un modèle générique de modélisation d'industrie de fonderie sable au travers du logiciel *Cost Advantage*. La première étape est de modéliser le processus de production propre à cette industrie. Il s'en suit une détermination des paramètres influant sur le coût de production d'une pièce pour enrichir la sémantique coût du modèle.

Il s'est tout de suite posé le problème de l'approche générique. Comment définir des contextes et des caractéristiques qui permettent d'induire un coût pertinent de pièce à produire et avec quel niveau de détail pour parvenir à créer un modèle applicable à un grand nombre d'entreprises ?

A cette question nous n'apporterons pas de réponse tranchée mais nous éclairerons des pistes ou solutions que nous avons mises en œuvre.

Mais commençons par la présentation de l'organisation générale d'une fonderie sable en regardant le cycle de vie d'une pièce. Cette modélisation s'inspire de la fonderie acier SMC Colombier Fontaine (France) du groupe AFE Métal précédemment sitée.

Concernant la modélisation des coûts, nous restreindrons notre étude à la réalisation d'une pièce brute dont le cycle de vie en entreprise est présenté en Figure 5. Nous nous concentrerons pour cette première approche de modélisation aux aspects de production, de l'élaboration du sable à moule, des outillages, jusqu'au parachèvement des pièces.

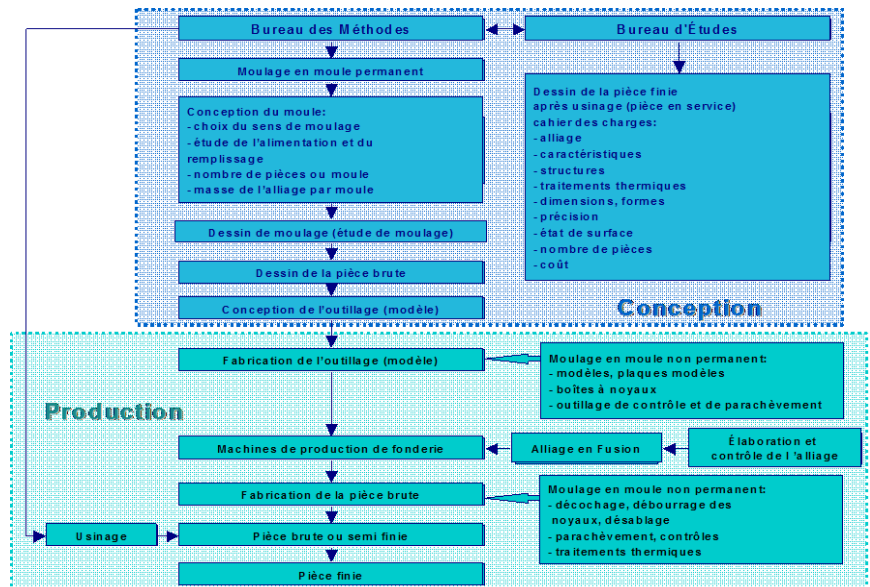

**Figure 5.** *Cycle de conception-production en fonderie sable (vue fonctionnelle)*

*4.3. Démarche induite par l'utilisation de Cost Advantage*

La Figure 6 présente une vue fonctionnelle modélisant en entrée les composants (matière première, outillages...) et les éléments nécessaires à la réalisation d'une pièce, ainsi que les principaux facteurs (rebus, pertes) affectant les calculs des coûts des pièces.

**Fig.6 : Cycle de production de pièces en fonderie sable (vue processus)**

**Fig.7 : Modélisation de la démarche d'obtention d'une pièce brute sous Cost Advantage au niveau de l'assemblage**

En correspondance, la Figure 7 représente une transposition sous les concepts de *Cost Advantage* de cette modélisation regroupant les trois niveaux d'entités définies dans le logiciel. Pour illustrer cela, par exemple, le *moule* et les *noyaux* sont des composants nécessaires pour réaliser l'assemblage nommé *moulage* par l'opération (feature) de *remoulage*.

Il est donc nécessaire, pour définir la pièce brut, d'effectuer les deux assemblages que sont dans un premier temps le *moulage* (réalisation du moule) puis la *coulée*. Avec pour chaque assemblage comme indiqué précédemment la nécessité de renseigner l'ensemble des informations portant soit sur les composants, soit sur les opérations.

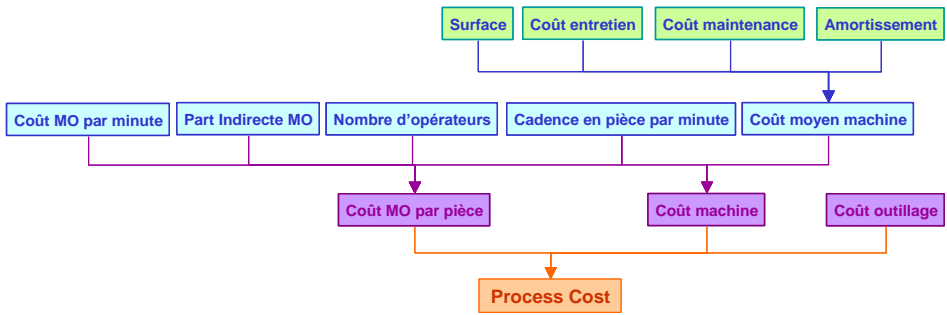

**Fig.8 : Description de la structure du coût d'un procédé, d'une opération**

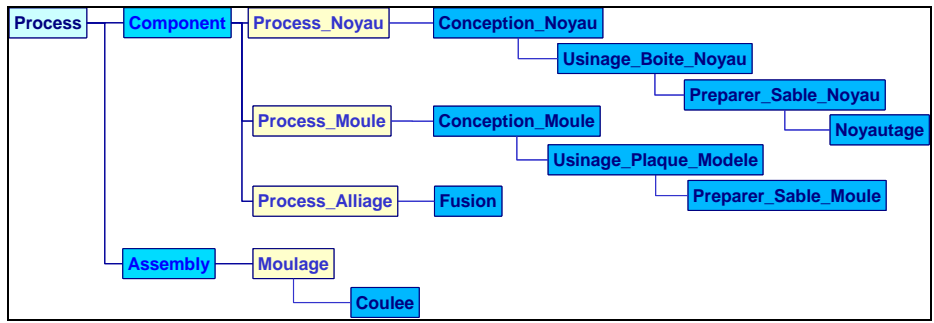

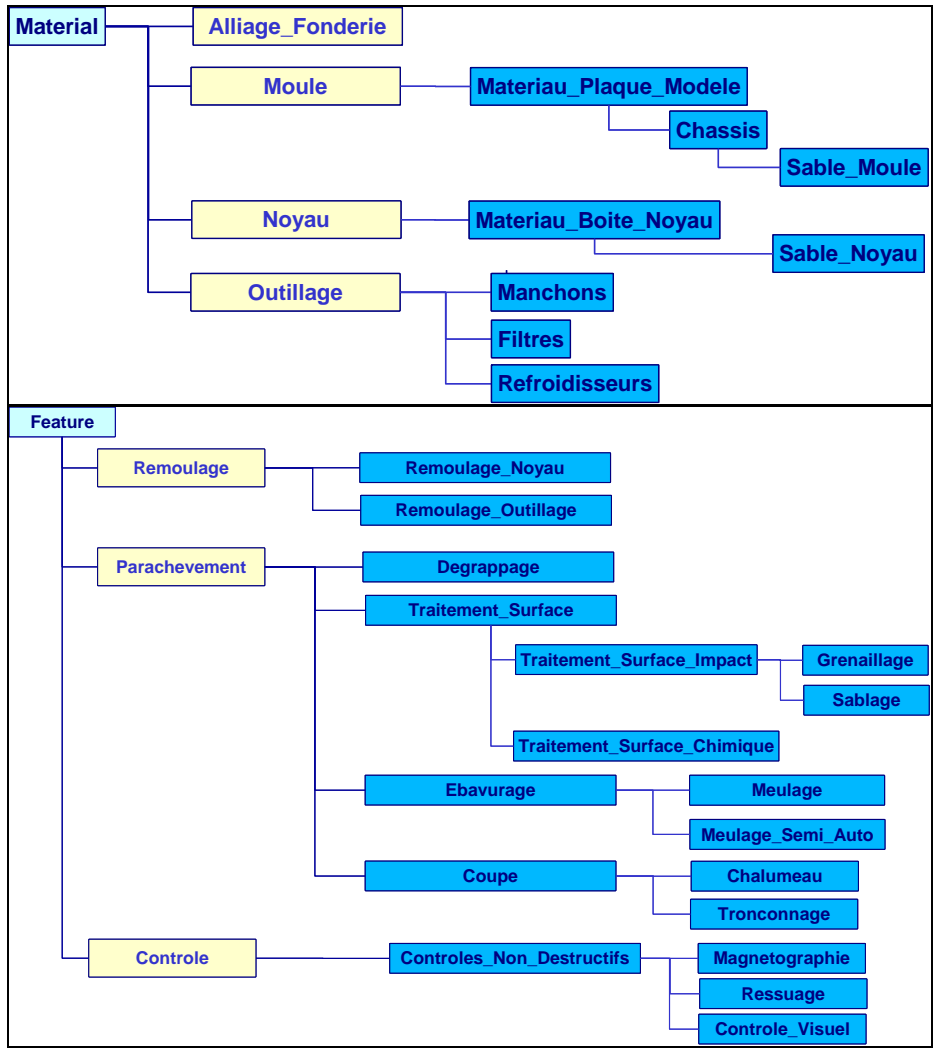

**Fig.9 : Structure de données sous *Cost Advantage***

En terme de conception de modèle, la vue fonctionnelle identifie les assemblages à faire, il faut alors définir les composants nécessaires ainsi que statuer sur les opérations à définir. Il est nécessaire dans l'outil d'avoir une démarche ascendante, à savoir, partir des composants pour remonter jusqu'à la définition des assemblages.

Le calcul des coûts de pièce est repris dans la figure 8 et la structure des donnée implémentée en figure 9. Les calculs sont simples traduisant des volumes de matière, les taux de pertes et les coûts de main d'œuvre et de machine nécessaires. Mise à part la difficulté de connaître les paramètres exacts, les règles de calculs sont de simples additions et multiplications ; là n'est pas le problème de modélisation qui se positionne plus sur une problématique de hiérarchisation et organisation du modèle.

Les règles de calcul alors implémentées permettront au futur utilisateur de ne renseigner que les données pertinentes sur son étude. En effet, seul le processus opératoire, cadences, dimensions, nombres de noyaux (etc..) seront demandées (ou déduits directement depuis un logiciel CAO) pour permettre un calcul automatique du coût de la pièce en fonction de ses caractéristiques propres.

## 5. Discussion

La principale remarque que l'on peut faire sur le travail développé, vient du fait qu'il n'a pas intégré l'aspect global de la gestion des coûts. A savoir, la part indirecte due au développement et à la conception des outillages (plaque modèle, boite à noyaux…). Ceci réduit fortement l'application de cet outil à un générateur de devis. Il est nécessaire de pouvoir connaître pour ce faire, la participation exacte des parties indirectes et donc la quantité globale commandée pour connaître l'incidence sur chaque pièce ou sur la quantité globale.

Mais gardons en tête le cadre d'utilisation d'un tel outil : aider le concepteur à atteindre un objectif coût lié à la pièce qu'il conçoit. Il peut entre autre proposer des modifications de conception (plan de joint et donc noyaux, qualité attendue…) ou de process (nombre de pièces au moule par exemple…) pour atteindre son objectif. De plus, quand l'ensemble des partenaires a validé cette conception, les paramètres coûts sont figés en accord ou non avec l'objectif fixé initialement, l'indicateur coût est figé et ne sera plus une conséquence de décisions ultérieures.

Nous avons au cours de ce travail identifié une difficulté principale qui pour cette modélisation réside dans la multiplicité des éléments à caractériser et dans leur hiérarchisation. Même si le processus de fabrication d'une pièce en fonderie sable semble simple, il fait intervenir de nombreux composants (alliage, noyaux, moule…) dont nous avons volontairement limité la définition en terme de raffinement de modèle puisque chacun de ces composants peut faire l'objet d'une modélisation plus fine. Cette limitation nous l'avons choisie afin de représenter au mieux le processus général sans s'approcher trop des spécificités métiers propres à chaque entreprise.

Par exemple, le processus de réalisation des noyaux ou de fusion n'est pas complètement défini puisque paramétrable en fonction des machines, usages et autres spécificités de l'atelier de fusion ou de noyautage. En fait cette remarque prévaut pour tous les composants entrant dans la réalisation de la pièce finie, mais nous pensons avoir défini un squelette de base minimal, transposable d'une entreprise à l'autre utilisant le même processus à savoir la fonderie sable. Il semble relativement aisé alors de définir un modèle propre à la fonderie coquille.

Or cette dernière remarque nous rappelle que nous avons peu pris en compte la part indirecte (étude et conception de la grappe et des outillages) puisqu'il est actuellement difficile d'affecter des indicateurs pertinents et fiables aux temps liés à l'étude. Nous nous sommes concentrés sur les aspects liés à la réalisation de la pièce. Le modèle proposé, que nous avons volontairement limité en terme de détail, permet d'obtenir une quantification des coûts d'une pièce intégrant les cadences, rendements, taux de perte etc.

## 6. Conclusion

Pour conclure sur ce travail, nous avons commencé à appréhender une logique de modélisation orientée coût au travers d'un outil utilisant le concept d'entité coût. En vue d'assurer un aspect générique à notre travail, nous avons délibérément limité les détails des opérations, composants et assemblages (principalement pour l'obtention des noyaux ou du sable). En effet, l'élaboration de ces éléments prend en compte de nombreux paramètres (sable, recyclage, éléments d'addition, technique de réalisation des noyaux, etc..) qu'il nous a semblé surabondant dans un premier temps de définir.

Nous avons ainsi défini une structure que nous pensons minimale, ainsi que des indicateurs nécessaires pour évaluer l'ensemble des coûts abordés. La suite à donner est, entre autre, d'appliquer cette modélisation à différentes entreprises du métier de la fonderie sable en configurant le modèle aux processus existants et renseignant les valeurs exactes des indicateurs définis (taux horaires, rendements, cadences etc..). Cette dernière étape permettra de comparer l'efficacité des différentes entreprises et pourra servir de Benchmark. Une difficulté prévisible, concerne la possibilité d'accéder à ces informations. De plus, ces facteurs sont souvent gérés par une comptabilité analytique globale donc noyés au sein d'indicateurs et de systèmes de gestion peu transparent ou affectés aléatoirement.

L'autre suite importante à donner à ce travail est la prise en compte globale des coûts liés principalement aux parts indirectes. Notre introduction met en exergue le manque de gestion de ces aspects et notre première approche n'a pas donné lieu à une meilleure maîtrise de ces facteurs. Pourtant nous sommes convaincus de la nécessité de maîtriser ces aspects. La première étape a consisté à traduire le processus de réalisation des pièces ce qui réduit l'utilisation à la génération d'un coût de pièce usine. Une meilleure spécification (par le biais d'indicateurs, de métriques) des phases de conception d'outillages, durée de vie des outillages, etc.. permettrait d'intégrer un coût réel non plus d'une pièce, mais de la série complète. Se pose alors la question de la pertinence de l'outil utilisé pour ce type d'approche.

## 7. Bibliographie

Perrin J., « Cohérence, pertinence et évaluation économique des activités de conception », in *Cohérence, Pertinence et Evaluation*, ECOSIP, Economica, 1996.

Bouin X., Simon F-X., « Les nouveaux visages du contrôle de gestion », *Approches Techniques et Comportementales* - Paris: DUNOD, ISBN 2 10004887 2, 2000.

Milkoff R. « Le concept de comptabilité à base d'activités » *Rapport de recherche GREGOR*, IAE Paris 1996 – http ://www.univ-paris1.fr/GREGOR/.

Bernard A., Delplace J.-C., Perry N., « Cost Analysis as the base of product realization cycle », *CIPR General Assembly annals 2003 on Manufacturing technology*, ISBN 3 905 277 39 5, 2003.

Brodier J.P., « La dictature des prix de revient » Harvard - L'Expansion, 1990, www.directva.com/PDF/Dicta2.pdf

Alcouffe C., Bes M-P., « Standardisation et évaluation des activités de conception dans le secteur aérospatial », *Deuxième Congrès International Franco-Québécois de Génie Industriel*, ALBI, France, 1997.

Bellut S., « Estimer le coût d'un projet », Saint Denis de La Plaine : AFNOR Gestion, Edition 2002.

Mevellec P. « Le coût global, nouvelle frontière du calcul de coûts », FINECO (en attente de publication), 2001.

Delafollie G., « Analyse de la Valeur », Paris: Hachette Technique, ISBN 2 01017151 9, 1991.

Petitdemange C., « La maîtrise de la valeur » *La gestion de projet et l'ingénierie simultanée* – Paris-La Défense: Afnor Gestion, ISBN 2 12475021 6, 1991.

Giannopoulos N., Roy R., Taratoukhine V., Griggs T., « Embedded systems software cost estimating within the concurrent engineering environment », *CE : the Vision for the future generation in research and applications*, ISBN 90 5809 622 X, 2003, p.353-357.

Kieffer J-P., Pujo P., « Méthodes du pilotage des systèmes de production », Hermès Sciences Lavoisier, ISBN 2 74620514 9, 2002.

Bernard A., Perry N., Delplace, J.-C., Gabriel S., "Optimization of complete design process for sand casting foundry", *Proceedings of IDMME'2002 conference*, Clermont-Ferrand, France, 2002.

Perry N., Bernard A., "Cost objective PLM and CE", *CE : the Vision for the future generation in research and applications*, ISBN 90 5809 622 X, 2003, p.817-822.

H'Mida F., « L'approche entité coût pour l'estimation des coûts en production mécanique », Thèse de Doctorat, LGIPM-ENSAM Metz-Univ. Metz, Mars 2002.